\documentclass[reprint,amsmath,amssymb,aps,pra]{revtex4-1}
\usepackage[utf8]{inputenc}
\usepackage{graphicx}
\usepackage{dcolumn}
\usepackage{mathrsfs,amsmath}
\usepackage{bm}
\usepackage{xcolor}
\def\ph2{{\it p}-H$_2$}
\def\Am3{\AA$^{-3}$}

\renewcommand{\vec}[1]{\mathbf{#1}}
\begin{document}
\title{Roton excitation in overpressurized superfluid $^4$He}
\author{Youssef Kora and Massimo Boninsegni}
\affiliation{Department of Physics, University of Alberta, Edmonton, Alberta, T6G 2E1, Canada}
\date{\today}

\begin{abstract}
    We carry out a theoretical investigation of overpressurized superfluid phases of $^4$He  by means of quantum Monte Carlo (QMC) simulations. As a function of density, we study structural and superfluid properties, and estimate the energy of the roton excitation by inverting imaginary-time density correlation functions computed by QMC, using Maximum Entropy. We estimate the pressure at which the roton energy vanishes to be about 100 bars, which we identify with the spinodal density,  i.e., the upper limit for the existence of a metastable superfluid phase.
\end{abstract}
\maketitle

\section{Introduction}
Helium is the only element in nature that does not crystallize at zero temperature under the pressure of its own vapor; instead, its thermodynamic equilibrium phase is a liquid capable of flowing without dissipation ({\em superfluid}). A pressure of around 25 bars must be applied  in order to stabilize a hexagonal closed-packed crystalline phase. There is now consensus that, upon crystallizing, the system loses its superfluid properties \cite{colloquium}.
\\ \indent
 It is possible, however, to realize experimentally metastable liquid phases of helium at  pressures higher than that of crystallization \cite{balibar, werner}. This allows one to study the the superfluid response of the system over a significantly greater range of pressure. 
 Theoretical studies have shown that at temperature $T=0$ the condensate fraction remains finite in the overpressurized liquid, decaying exponentially with density \cite{moroni2004}. Computer simulations have also yielded evidence of a possible ``superglass'' phase, with an estimated lifetime of the order of 1 ms, displaying a finite superfluid response but also breaking translational invariance over relatively long time scales \cite{superglass}. 
The predicted resilience  of the overpressurized superfluid phase of $^4$He is understood to be a direct consequence of quantum-mechanical exchanges involving a macroscopic fraction of all particles in the system (an effect also referred to as "quantum jamming") \cite{jamming}.  
\\ \indent
Of particular interest is whether superfluidity persists all the way to the limit of existence of a metastable overpressurized fluid. This limit is identified by a value of density, henceforth referred to as {\em spinodal}, above which only the crystalline phase occurs. It is speculated that the energy of the minimum of the excitation spectrum of superfluid $^4$He at finite wave vector, known as the {\em roton}, should vanish at the spinodal density \cite{nozieres}.
\\ \indent
The roton energy as a function of pressure has been measured experimentally in the equilibrium fluid phase up to a pressure of 20 bars \cite{glyde,andersen1994,gibbs}, as well as in various porous media, in which the fluid phase can be stabilized above the bulk freezing pressure, as crystallization is suppressed by the tight confinement \cite{molz}. The highest pressure at which superfluidity has been observed in porous media is $\sim 37$ bars, where the roton mode disappears \cite{glyde08,glyde18}.
However, no measurement of the roton energy in the overpressurized bulk superfluid, which has been predicted to exist at much higher pressures, has to our knowledge been carried out yet. 
\\ \indent
Besides the outstanding theoretical issue mentioned above, namely the behavior of the roton energy on approaching the spinodal,  the parallel  behavior of the superfluid and condensate fraction at finite temperature, as a function of pressure, is also of interest; there exist ground state studies of the condensate fraction of overpressurized superfluid $^4$He, but it is known that the superfluid fraction must be equal to 100\% in the ground state of a translationally invariant system.
Furthermore,
since the excitation spectrum can be probed by neutron scattering measurements, knowledge of the roton energy as a function of density and pressure can be used to gain information about the local environment experienced by the fluid in confinement or in restricted geometries.
\\ \indent
We report here results of a theoretical investigation of overpressurized superfluid $^4$He, carried out by means of first principle QMC simulations at temperature $T=1$ K.  
The goal of this QMC study is to examine the structural and superfluid properties of the metastable superfluid  phase at very high pressures, as well as to calculate the energy associated to the roton minimum of the elementary excitation spectrum. This task is complicated by the lack of a direct probe of real-time dynamical properties of the system within the context of QMC. There are, however, indirect ways of extracting some of that information, such as computing imaginary-time correlation functions, and converting them to real-frequency spectral functions through an inverse Laplace transform. This is an ill-posed problem that requires the use of a regularization scheme; we use the Maximum Entropy Method (MEM) \cite{bayesian}. 
\\ \indent
Our main result is that the energy of the roton excitation vanishes at a density $\rho_{sp} =0.0320(2)$ \AA$^{-3}$. This is also the highest density for which the  simulation of a metastable, overpressurized superfluid phase of $^4$He is feasible, as spontaneous crystallization rapidly occurs at higher density, not allowing us to collect meaningful statistics. We can therefore identify $\rho_{sp}$ with the spinodal density, in agreement with the hypothesis of Ref. \onlinecite{nozieres}. The pressure corresponding to $\rho_{sp}$ is equal to 104 bars, to be compared to that (67 bars) of the equilibrium crystalline phase at the same density.
\\ \indent
We report estimates for the condensate fraction $n_0$ as a function of the density, and we find them to be in quantitative agreement with previous ground state studies, up to a pressure of approximately 60 bars; significant deviations are observed from the previously predicted exponential decay, at higher pressure, i.e., the condensate fraction decays considerably more rapidly with density. Analogously, the computed superfluid fraction $\rho_S$ remains relatively close to 100\% as te density is increased, but falls off abruptly on approaching $\rho_{sp}$.

The remainder of this paper is organized as follows: in section \ref{mo} we describe the model of the system, and briefly describe the regularization procedure we use to extract some dynamical properties of the system; in Sec. \ref{me} we  describe our QMC methodology; we present and discuss our results in Sec. \ref{res} and finally outline our conclusions in Sec. \ref{conc}.
\\ \indent

\section{Model}\label{mo}
We model the system as an ensemble of $N$ point-like, identical particles with  mass $m$ equal to that of a $^4$He atom and with spin $S=0$, thus obeying Bose statistics. The system is enclosed in a cubic cell of volume $V$ with periodic boundary conditions in the three directions. The density of the system is $\rho=N/V$. 
The quantum-mechanical many-body Hamiltonian reads as follows:
\begin{eqnarray}\label{u}
\hat H = - \lambda \sum_{i}\nabla^2_{i}+\sum_{i<j}v(r_{ij})
\end{eqnarray}
where the first (second) sum runs over all particles (pairs of particles), $\lambda\equiv\hbar^2/2m=6.06$ K\AA$^{2}$, $r_{ij}\equiv |{\bf r}_i-{\bf r}_j|$ and $v(r)$ denotes the pairwise interaction between the helium atoms. In this investigation, we model this interaction using the well-established Aziz pair potential \cite{aziz79}, which is the canonical model utilized in most numerical studies of superfluid helium. This model only includes pair-wise interactions; in principle additional terms should incorporated, describing non-additive energy contributions, the leading one involving triplets of atoms. However,
there are strong indications in the literature  that the effects of three-body corrections are negligible as far as the the structure and dynamics of the system are concerned. Their effect on the pressure, on the other hand, is believed to be no larger than 1-2\% in the range of densities considered in this work \cite{mpfb,ceperley94}.

\section{Methodology}\label{me}
As mentioned above, we carried out QMC simulations of the system described by Eq. (\ref{u}), using the continuous-space Worm Algorithm \cite{worm,worm2}.  This technique is by now well-established, and extensively described in the literature. We shall therefore not review it here, instead referring the reader to the original references. We utilized a canonical variant of the algorithm in which the total number of particles $N$ is held constant, in order to simulate the system at fixed density \cite{mezz1,mezz2}. 
\\ \indent 
The details of the QMC simulation are standard; we adopted the usual the short-time approximation for the imaginary-time propagator accurate to fourth order in the time step $\epsilon$ (see, for instance, Ref. \onlinecite{jltp}). All of the results presented here are extrapolated to the limit of vanishing $\epsilon$. The numerical estimates of the quantities of interest computed with $\epsilon=1.6\times 10^{-3}$ K$^{-1}$ are indistinguishable from the extrapolated ones, within the statistical uncertainties of the calculation. The results shown here were obtained for systems comprising $N=256$ particles. Experience with previous work \cite{massimo1996} suggests that this system size is sufficient to extract information at the roton wave vector, of interest here.
\\ \indent 
All calculations were carried out at $T=1$ K. For densities up to freezing, such a value of the temperature is well below the superfluid transition temperature $T_c$, and therefore our physical estimates may be expected to approach closely ground state values. For example, the excitation spectrum of the system is experimentally observed to be essentially independent of temperature, in this range of density (see, for instance, Refs. \onlinecite{gibbs,dietrich1972}). On the other hand, at higher density, in the overpressurized metastable regime, this is no longer guaranteed, as pressurization is expected to suppress $T_c$ (there are no experimental data nor theoretical estimates of which we are aware).
\\ \indent
The properties of the system are studied as a function of the density; below the freezing density $\rho_f$, equal to $\sim$ 0.0262 \AA$^{-3}$, simulations are straightforward, as one is studying the thermodynamic equilibrium phase. On the other hand, above the freezing and melting density ($\rho_m\sim 0.0286$ \AA$^{-3}$), the system starts displaying a marked tendency to crystallize, and an appropriate simulation protocol has to be adopted in order to prevent that from happening too quickly, in order to accumulate enough statistics for the metastable, homogeneous superfluid phase.
We adopted the same protocol as in Ref. \onlinecite{search}, i.e., we increase the density of the system in steps, by rescaling all particle coordinates (i.e., along imaginary-time world lines \cite{worm2}) in a many-particle configuration coming from a simulation at a slightly lower density (the immediately previous step). 
\\ \indent
The advantage of this approach is that one is starting from configurations that are already ``entangled'', i.e., they feature permutations of large numbers of particles. In order to reach the crystalline, equilibrium phase, the simulation algorithm must ``disentangle'' all of these world lines, and although this will of course eventually happen, the metastable phase may be sufficiently ``long-lived'' (in the computer) that  one may still arrive at physically meaningful  expectation values. Of course, there will always be a drift in the averages over the course of the simulation, as the true equilibrium phase inevitably emerges, but in most cases it is small enough not to be a concern.
\\ \indent 
In order to study the elementary excitations of the system, one can estimate the dynamic structure factor $S({\bf q},\omega)$, by calculating by QMC  the imaginary-time correlation function
\begin{equation}\label{propagator}
F(\vec{q},\tau)=\frac{1}{N}\ \langle\hat\rho_{\vec{q}}(\tau)\ \hat\rho_{\vec{q}}^\dagger(0)\rangle 
\end{equation}
where $\langle ...\rangle$ stands for thermal average, and with
\begin{align}\label{densityq}
\rho_{\vec{q}}({\tau}) = \sum_{j=1}^N  e^{i \vec{q} \cdot \vec{r}_{j}},
\end{align}
where the $\{{\bf r}_j\}$, $j=1,2,...N$ are the positions of the $N$ $^4$He atoms at imaginary time $\tau$, and inferring $S({\bf q},\omega)$
through 
\begin{align}\label{fourier2}
F(\vec{q},\tau) = \int_0^{\infty}\ d\omega\ (e^{-\omega \tau}+
e^{-\omega(\beta-\tau)})\ S(\vec{q},\omega)
\end{align}
where $\beta=1/T$, $0\le\tau\le\beta$
(we have set the physical constants $\hbar=k_B=1$). 
As mentioned above, the inversion in (\ref{fourier2}) constitutes a mathematically ill-posed problem, and we use the MEM to obtain the position of the main peak of $S({\bf q},\omega)$ (i.e., the energy of the excitation dominating the spectrum) as a function of density. 
\\ \indent
The MEM (and closely related approaches) has been adopted in the past to estimate the dynamic structure factor of superfluid and normal $^4$He \cite{massimo1996,gift,maxent}; in general,  while the sharpest features of the underlying image tend to be lost in the reconstruction, usually the position of the main peak is rather accurately identified. 
\begin{figure}[h]
\centering
\includegraphics[width=0.40\textwidth]{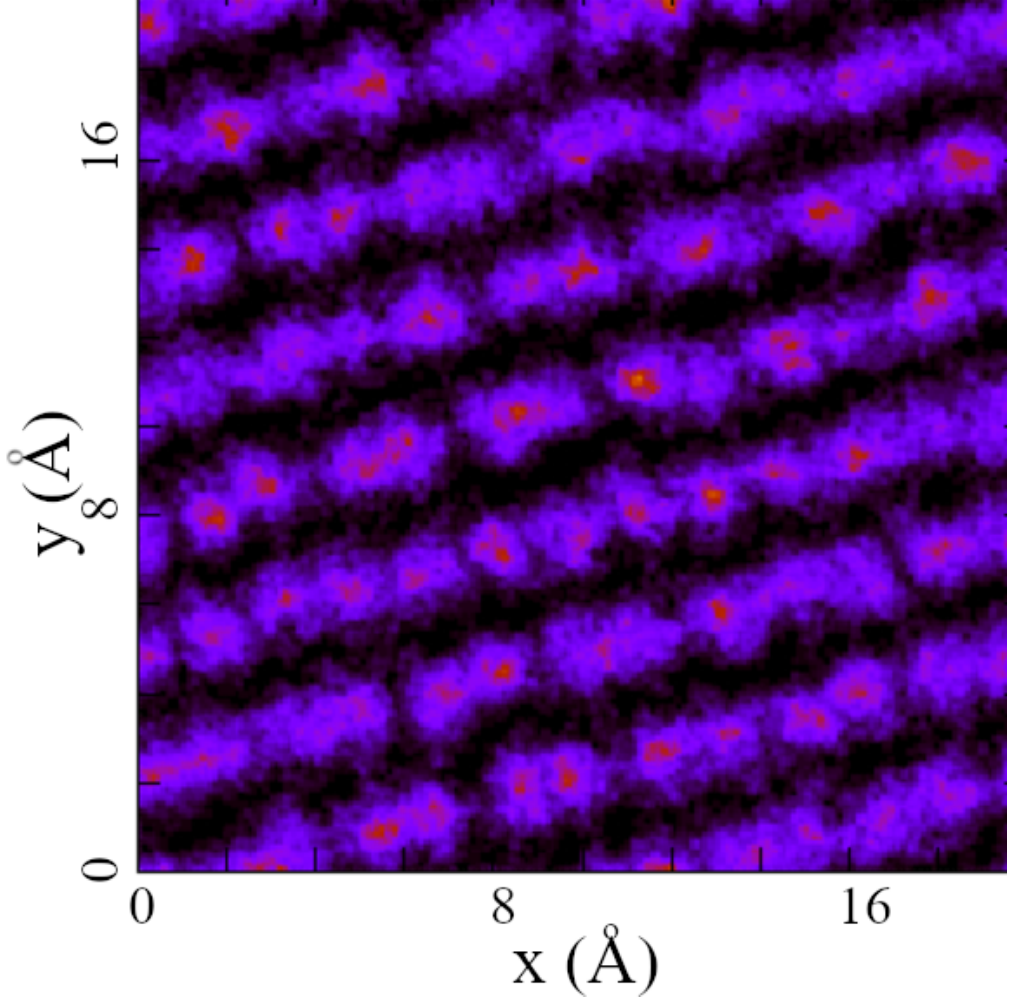} 
\caption{{\em Color online}. Instantaneous density map of a system of $N=256$ $^4$He atoms (view is along the $z$ direction) in a cubic box, at $T=1$ K and density 0.0336 \AA$^{-3}$. Clearly, in this case the system has crystallized.}
\label{config}
\end{figure}
\begin{figure}[h]
\centering
\includegraphics[width=0.47\textwidth]{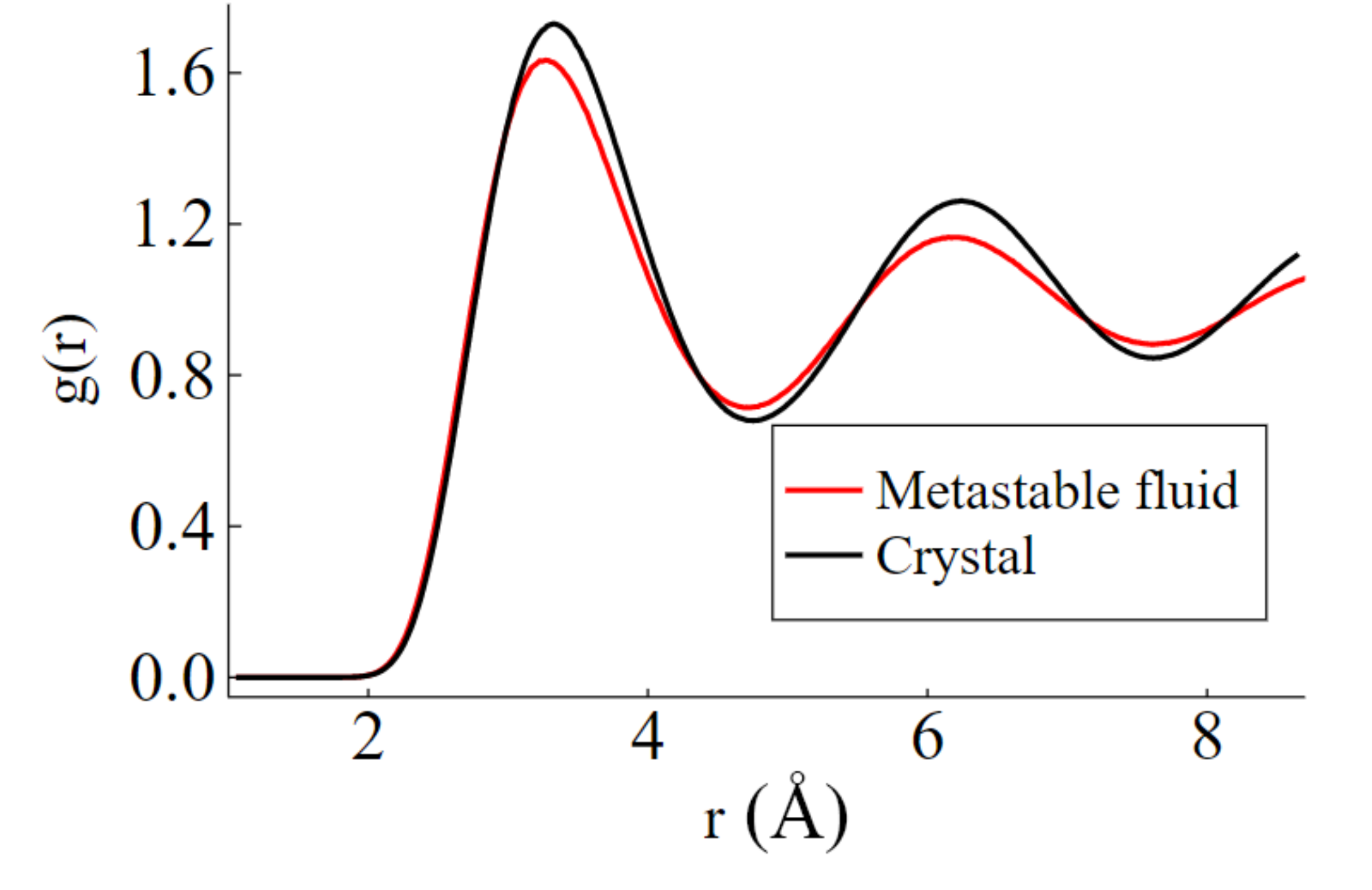} 
\caption{{\em Color online}. Pair correlation function of $^4$He at $T=1$ K and a density of $\rho=0.0319$ \AA$^{-3}$, for both the metastable superfluid and the equilibrium ({\em hcp}) crystalline phase (darker curve).}
\label{gr}
\end{figure}
In this work, we have not attempted the full reconstruction of the spectral image $S({\bf q},\omega)$ as a function of the wave vector ${\bf q}$, in order to obtain the energy dispersion curve $\omega({\bf q})$, thereby identifying the position of the roton minimum for each and every one of the densities considered. Rather, we have focused for simplicity on a single wave vector for each density, assuming that the magnitudes of the roton wave vectors $q$, $q^\prime$ at two different densities $\rho$ and $\rho^\prime$ be related through
$(q^\prime/q)=(\rho^\prime/\rho)^{1/3}$, as is experimentally found to be the case for the equilibrium superfluid phase below freezing \cite{pearce}.
\\ \indent 
As mentioned above, since we are using an equilibrium simulation technique, on simulating the system for a sufficiently long time eventually crystalline order is bound to emerge. It is therefore necessary to monitor the simulation in order to ensure that one is actually studying a metastable superfluid phase, and that crystal order has not yet set in. This is accomplished first and foremost by visual inspection of the many-particle configurations (i.e., imaginary-time paths) generated in the course of the simulation. As shown in Fig. \ref{config}, it is possible to detect the appearance of order rather easily, as it sets in even if the geometry of the box (cubic) is not specifically designed to accommodate a crystal of the known equilibrium structure ({\em  hcp} in the case of $^4$He). Another way to monitor the appearance of crystalline order is through the calculation of the pair correlation function, and the comparison with that (computed separately) of the equilibrium crystalline phase at the same density. An example of this is shown in Fig, \ref{gr}; although the two functions follow one another quite closely, that of the crystal has noticeably higher peaks.
\\ \indent
Another important indicator that one is simulating a metastable superfluid phase, besides of course the value of the superfluid fraction ($\rho_S$), which  is computed through the well-established winding number estimator \cite{pollock}, is  the one-body density matrix $n(r)$, which is expected to plateau at long distances in a superfluid, while decaying exponentially in a crystal.

\section{Results}\label{res}
In this section, we present our results for structural, superfluid, and dynamical properties of the overpressurized metastable phase of $^4$He.
\begin{figure}[h]
\centering
\includegraphics[width=0.47\textwidth]{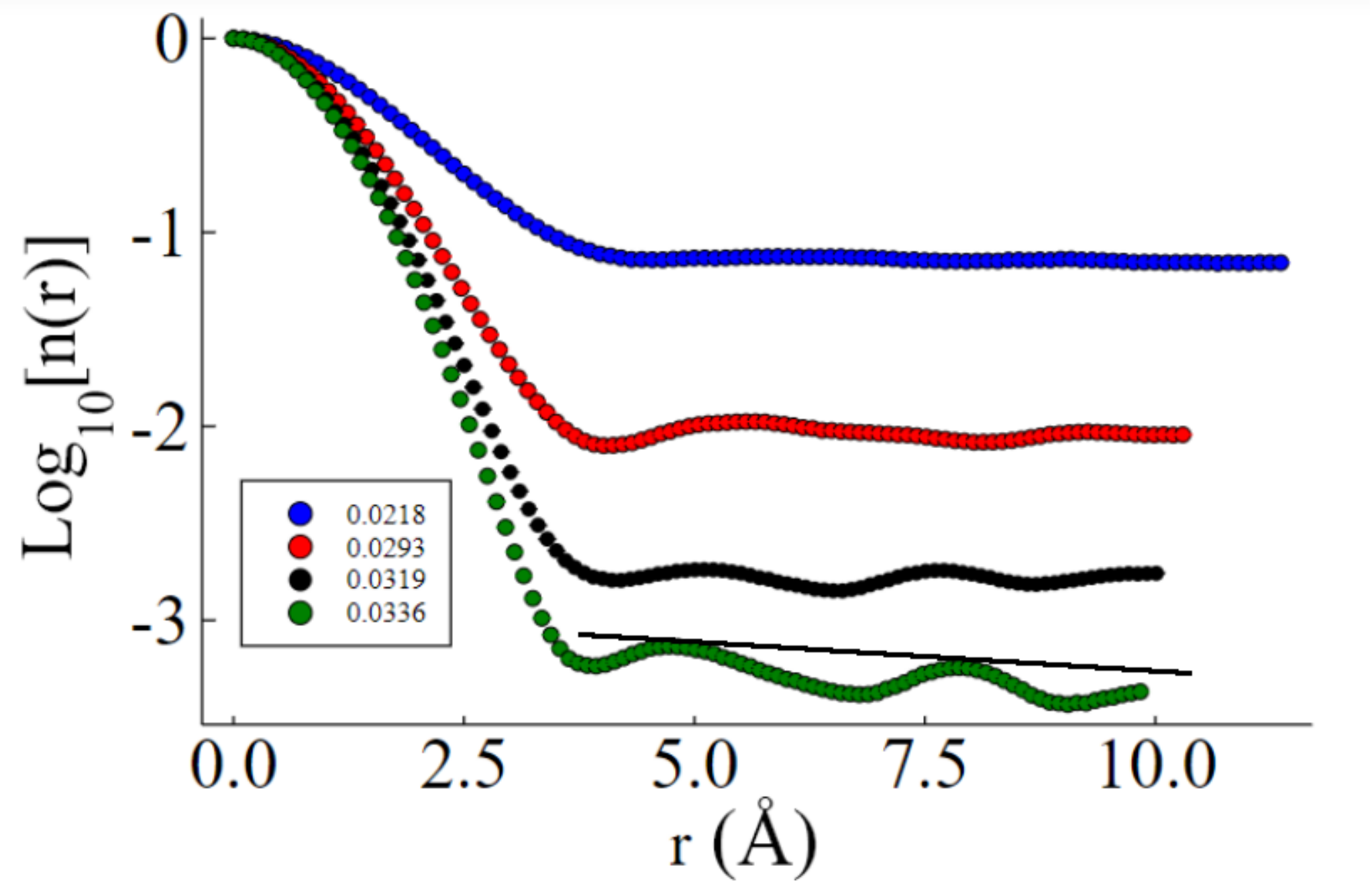} 
\caption{{\em Color online}. One-body density matrix of the metastable liquid phase of $^4$He at $T=1$ K and at various increasing densities (higher density shown by lower curve). The lowest density for which results are shown is the equilibrium density, the highest (bottom curve) 0.0336 \AA$^{-3}$. The straight line through the peaks of the bottom curve illustrates the consistency of the data with exponential decay.}
\label{nr}
\end{figure}

Fig. \ref{nr} shows the one-body density matrix $n(r)$ for a few different densities explored in this work. The lowest density for which results are shown is the equilibrium density , $\rho_{eq} = 0.021834$ \AA$^{-3}$. For the  highest density, namely 0.0336 \AA$^{-3}$,  data are  consistent with an exponential decay, suggesting that this density is above the spinodal. For all other densities, $n(r)$ plateaus at long distances to a value corresponding to  the condensate fraction  $n_0$. 

\begin{figure}[h]
\centering
\includegraphics[width=0.47\textwidth]{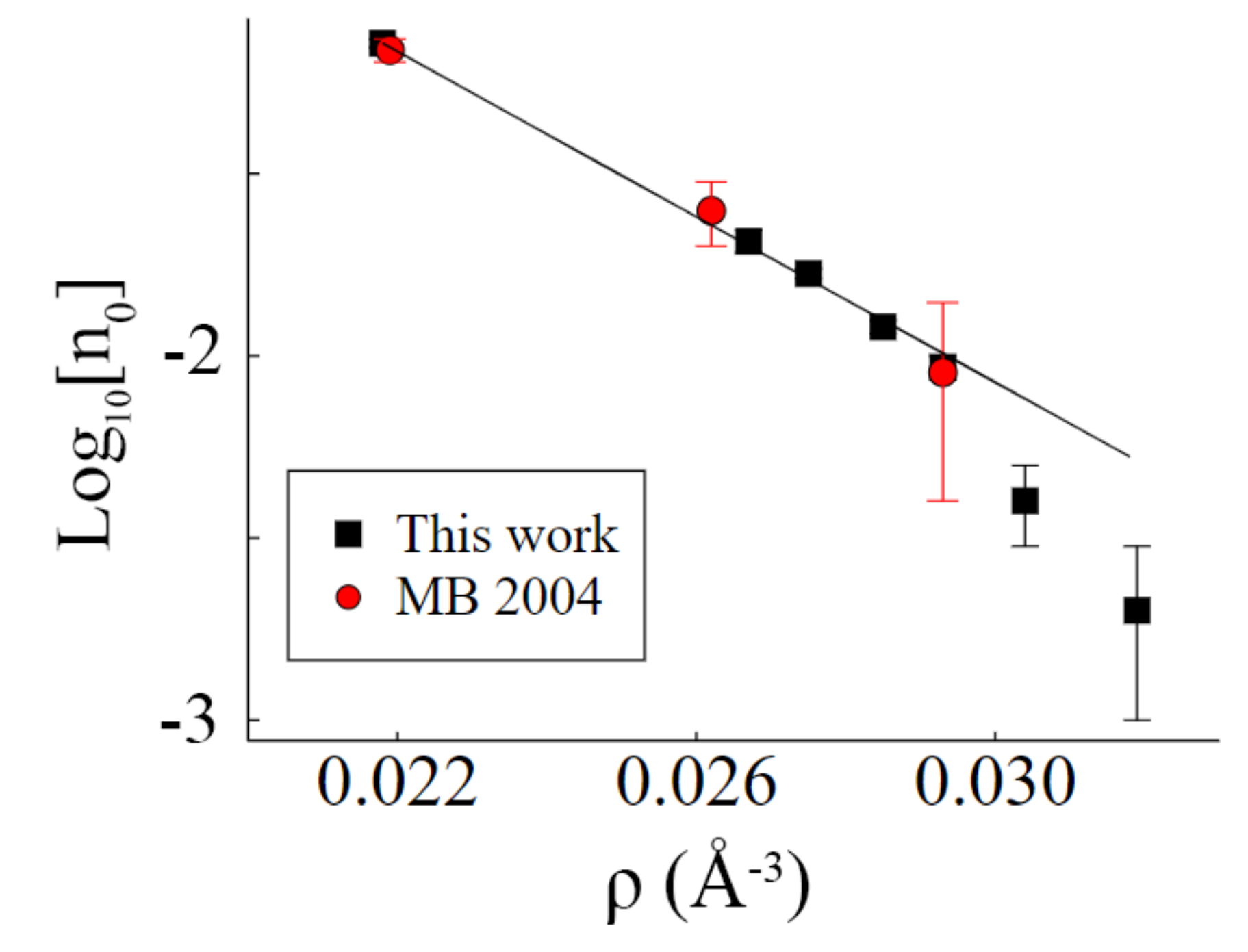} 
\caption{{\em Color online}. Condensate fraction ($n_0$) of the metastable liquid phase of $^4$He at $T=1$ K, as a function of density (squares). Also shown are the ground state estimates of  Ref. \cite{moroni2004} (circles).}
\label{nc}
\end{figure}
Fig. \ref{nc} shows our results for the condensate fraction as a function of density, comparing them with those for the ground state, obtained in Ref. \onlinecite{moroni2004}. The results of the two calculations are in perfect agreement, i.e., consistent with an exponential decay of the condensate fraction with density. However, in this work we considered densities $\sim 15$\% higher than in Ref. \onlinecite{moroni2004}; the data shown in Fig. \ref{nc} show significant deviations from the exponential decay, i.e., the condensate fraction decays more rapidly on approaching $\rho_{sp}$. Assuming that our statistical and systematic errors are not significantly underestimated (we believe this to be unlikely), one possibility to account for such deviations is that $T_c$ may be substantially suppressed, as the density approaches $\rho_{sp}$, and therefore the comparison of our results with ground state estimates may be complicated by thermal effects.

\begin{figure}[h]
\centering
\includegraphics[width=0.47\textwidth]{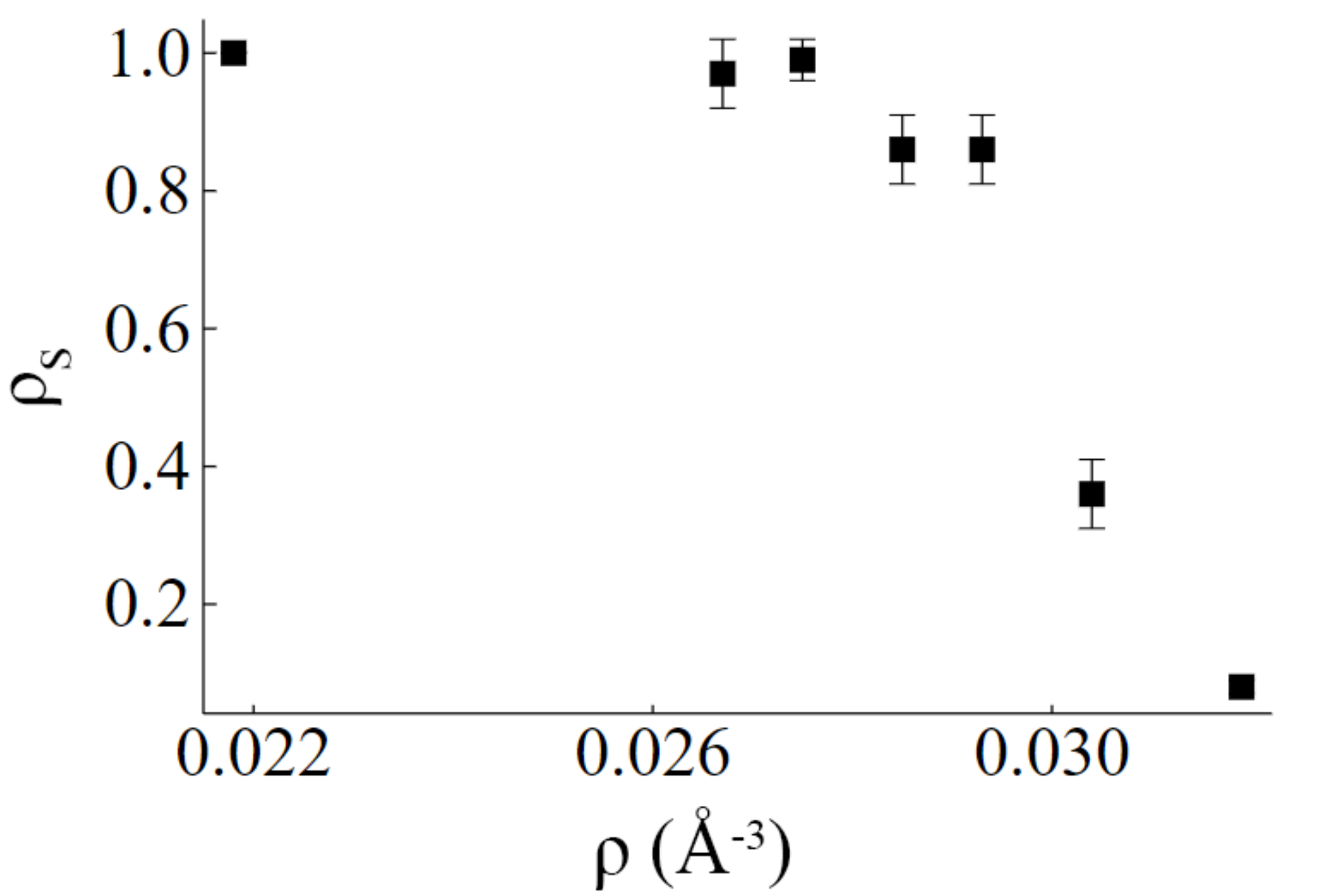} 
\caption{{\em Color online}. The superfluid fraction of the metastable fluid  phase of $^4$He at $T=1$ K, as a function of density.}
\label{rhos}
\end{figure}

\begin{figure}[h]
\centering
\includegraphics[width=0.47\textwidth]{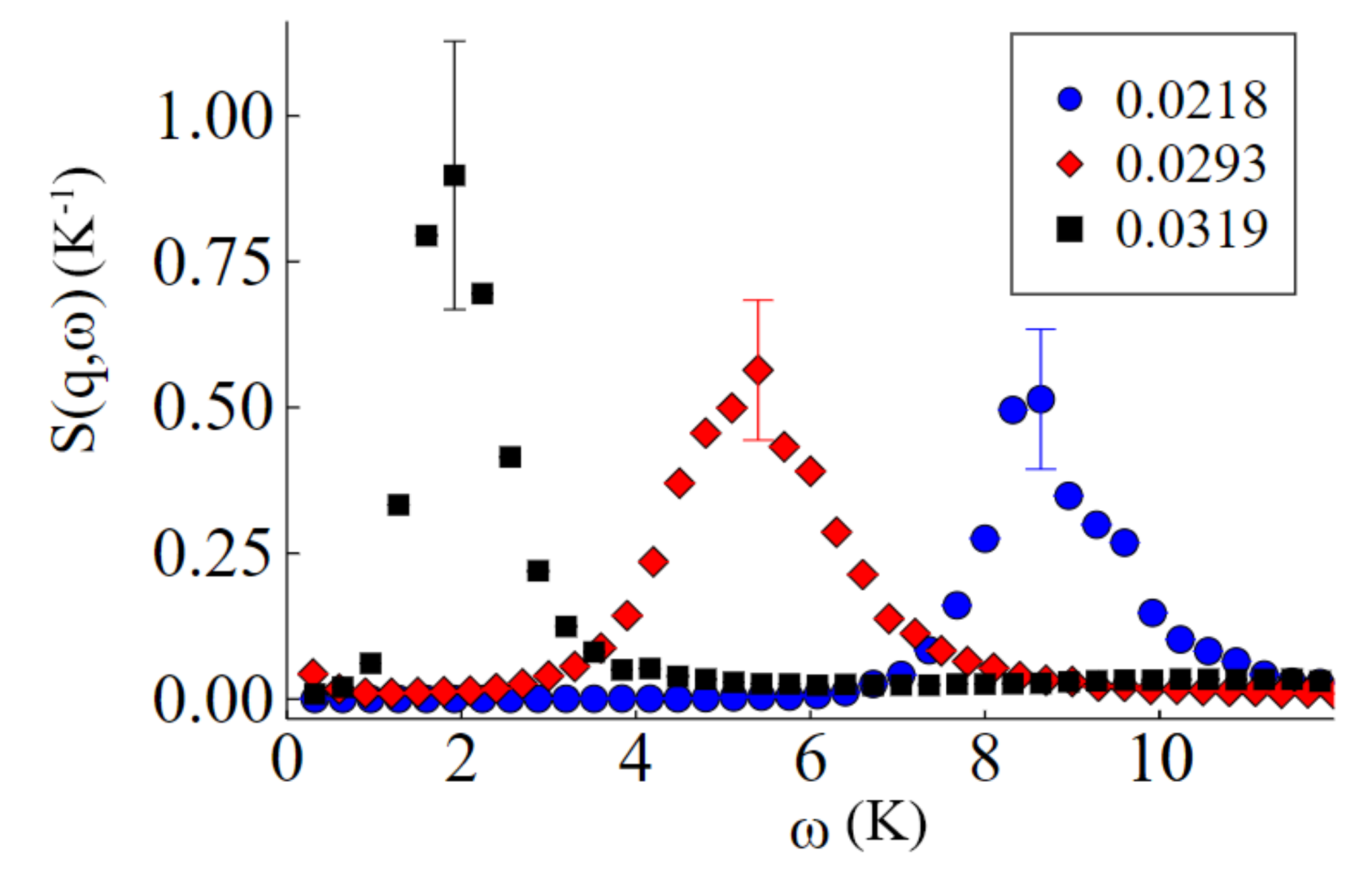} 
\caption{{\em Color online}. The dynamic structure factor $S(\vec{q},\omega)$ of superfluid $^4$He at $T=1$ K, evaluated  at densities of $\rho_{eq}$ ($q_R=1.963$ \AA$^{-1}$, circles), $\rho=0.0293$ \AA$^{-3}$ ($q_R=2.159$ \AA$^{-1}$, diamonds), and $\rho=0.0319$ \AA$^{-3}$ ($q_R=2.219$ \AA$^{-1}$, squares). The standard deviation associated with the inversion process is shown only for the peaks of the curves, with the understanding that the adjacent points have comparable or smaller standard deviations.}
\label{sqw}
\end{figure}

This hypothesis is corroborated by the values of the superfluid fraction, reported in Fig. \ref{rhos}. As one can see, while $\rho_S$ depends very weakly on $\rho$, remaining relatively large up to the highest density considered in Ref. \onlinecite{moroni2004} (0.0293 \AA$^{-3}$, corresponding to a pressure of approximately 60 bars), it decays abruptly above it, barely reaching $\sim 10$\% at the highest density for which a metastable superfluid phase can be simulated, using our protocol, namely 0.0319 \AA$^{-3}$.
\begin{table}[h]
 \begin{tabular}{|c | c |  c| c| c| c| c|} 
 \hline
  & \multicolumn{3}{c|}{Superfluid} & \multicolumn{1}{c|}{\em hcp}  \\
 \hline
 $\rho$ (\AA$^{-3}$) & $\rho_s$ & $n_0$ & P  & P  \\ [0.5ex] 
 \hline
 0.0293 & 0.86(5) & 0.0090(5)  & 62.0(3)  & 32.2(2)  \\ 
 0.0304 & 0.36(5) & 0.0040(5)  & 71.4(9)  & 45.2(3)   \\ 
 0.0319 & 0.08(1) & 0.0020(4) & 96(1) &  67.1(7)\\ [1ex] 
 \hline
 \end{tabular}
\caption{Superfluid ($\rho_s$) and condensate fraction ($n_0$), as well as the computed value of the pressure ($P$, in bars) for metastable superfluid $^4$He at $T=1$ K at different densities above the melting density. Statistical errors (in parentheses) are on the last digit. Also shown for comparison is the computed pressure for the equilibrium crystalline ({\em hcp}) phase.}
\label{table}
\end{table}
\\ \indent 
We report in Table \ref{table} values of the superfluid and condensate fraction, as well as computed pressure (in bars) for two different densities. Also shown for comparison are the values of the pressure for the corresponding equilibrium (crystalline {\em hcp}) phase, obtained separately in this work. As expected, the pressure is considerably higher for the metastable superfluid phase.
\\ \indent
Next, we discuss the results for $S(q_R,\omega)$, which constitute the most important part of this study ($q_R$ is the magnitude of the roton wave vector).
Fig. \ref{sqw} shows our results for $S(q_{R},\omega$), inferred through the MEM for the metastable superfluid phase at three different densities, including the equilibrium density $\rho_{eq}$ defined above. The results for the two higher densities are for two overpressurized superfluid phase.

All of the curves feature a well-defined maximum, whose position corresponds to the energy of the excitation. We estimate the position of the peak and assign a statistical uncertainty following the procedure outlined in Ref. \onlinecite{maxent}. Namely, we perform a Metropolis Monte Carlo simulation in the space of spectral images and accumulate statistics on the position of the maximum of the curve, also obtaining the uncertainty of its position as the standard deviation.
As expected, and as shown in Fig. \ref{sqw}, the roton energy goes down as a function of density. In addition, the height of the peak grows  as one approaches the spinodal density, and the onset of crystallization.

\begin{figure}[h]
\centering
\includegraphics[width=0.47\textwidth]{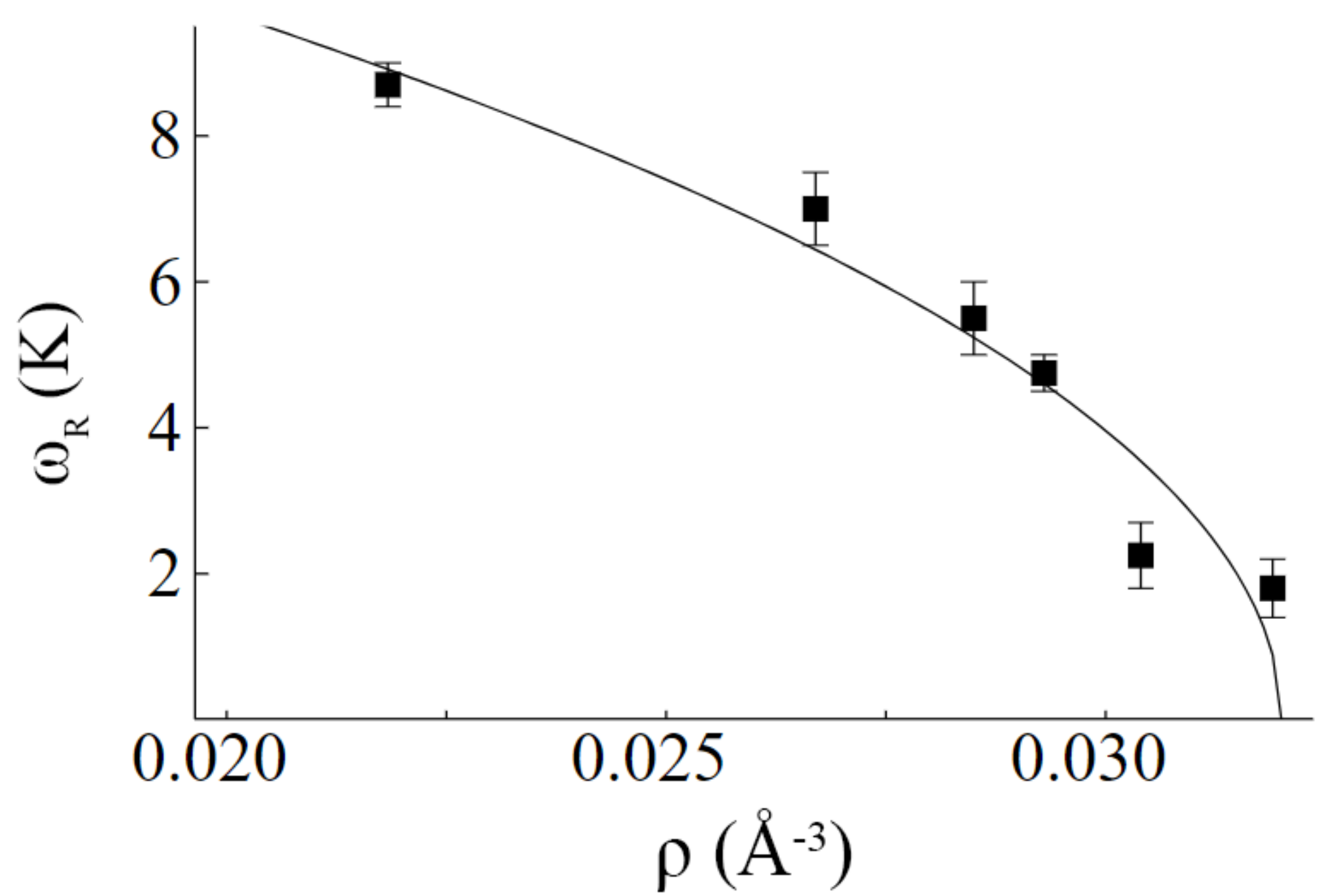}
\caption{The roton energy of superfluid $^4$He as a function of density, at $T=1$ K.}
\label{erot}
\end{figure} 

In Fig. \ref{erot}, we map out the roton energy  as a function of density, $\omega_R(\rho)$. In order to estimate the density at which $\omega_R=0$, we make the assumption that that occurs in concomitance with the divergence of the static structure factor, consistently with  Bijl-Feynman theory of the elementary excitations \cite{feynman}. This leads us to posit the following form \cite{nozieres}:
\begin{align}\label{fourier2}
\omega_R(\rho) = A(\rho_{sp}-\rho)^\gamma
\end{align}
We use this expression to fit the data in Fig. \ref{erot}, using $A$, $\rho_{sp}$ and the unknown exponent $\gamma$ as fitting parameters. This  yields   $\rho_{sp}=0.0320(2)$ \AA$^{-3}$, with a value of the critical exponent  $\gamma=0.12(5)$. This is consistent with the observed instability of the simulated fluid phase at $\rho=0.0336$ \AA$^{-3}$, and yields a value of approximately 100 bars for the upper limit to which the superfluid phase can be overpressurized.

\section{Conclusions}\label{conc}
We presented state-of-the-art QMC results for metastable superfluid phases of $^4$He, pressurized  above melting, at a temperature $T$= 1 K. These metastable phases can be rendered stable in a computer simulation (and presumably in Nature as well \cite{jamming,superglass}) by the presence of long cycles of exchange of $^4$He atoms, acting to prevent particles from becoming localized in space. This confers to the simulated metastable phase an appreciable ``lifetime'' (i.e., in the computer), that allows the meaningful measurement of physical observables. 
\\ \indent
The condensate fraction in the metastable overpressurized superfluid phase decays as a function of density, in a way that is consistent with the exponential decay predicted in previous ground state studies \cite{moroni2004}, up to a pressure of approximately 67 bars; concurrently, the superfluid fraction remains relatively close to 100\%. At higher pressures, not explored in previous calculations, we find that both the condensate and superfluid fractions decay more rapidly. This suggests that the superfluid transition temperature, relatively unaffected by pressure in the equilibrium superfluid phase, and even in the overpressurized phase for pressures below $\sim 67$ bars, becomes strongly suppressed at higher pressure.
\\ \indent
We computed the energy  of the roton excitation in the overpressurized superfluid phase, as a function of density. Our results are consistent with the hypothesis \cite{nozieres} that the roton energy should vanish at the spinodal density $\rho_{sp}$, in correspondence to a pressure of approximately 100 bars. Above such a pressure, an overpressurized superfluid phase is unstable against crystallization.
\\ \indent
The results of our study open up the possibility of more detailed experimental investigations of the overpressurized metastable liquid phases of helium, including in confined geometries (e.g., porous media). While the high pressures studied here are not necessarily directly measurable in an experimental setting, the roton energies that we compute are indeed measurable in the laboratory through neutron scattering techniques. The results we present here could therefore allow an indirect estimate of the local pressure of a metastable sample of overpressurized superfluid.

\section*{acknowledgments}
This work was supported by the Natural Sciences and Engineering Research Council of Canada. Computing support of Compute Canada is gratefully acknowledged. 

\bibliography{biblio}

\end{document}